# Mathematical Model of Hippocampal Microdialysis: Validation of *in vivo* Methodology

Damon Vinciguerra[a], Margot Vigeant[a] and Ewan McNay[b]

[a]Department of Chemical Engineering, Bucknell University, 701 Moore Avenue, Lewisburg, Pennsylvania, U.S.A.  dav006@bucknell.edu

[b]Behavioral Neuroscience, University at Albany, Albany, New York, U.S.A.

Please address proofs and reprint requests to Ewan McNay at: emcnay@albany.edu

Abbreviations used: ZNF, Zero-Net-Flux; ECF, Extracellular fluid; SEM, Scanning electron microscope; FEM, Finite element method

**Highlights**

- First-principles model accurately simulates in vivo results.
- Model allows investigation of variables not directly manipulable experimentally.
- Validity of zero-net-flux approach is supported.

# Abstract

Microdialysis is a well-established method for *in vivo* neurochemical measurements of small molecules, with implanted concentric-design probes offering minimized tissue damage and good temporal and spatial resolution. However, the large majority of measurements do not allow the perfusate to reach equilibrium with the brain, so that inferential methods of sample concentration correction such as zero-net-flux must be used to determine actual brain extracellular fluid glucose concentrations. In order for such methods to be valid, steady-state transfer of the analyte of interest within the brain is required, but this situation has not previously been confirmed. A first-principles mathematical model of fluid flow and analyte diffusion around an implanted microdialysis probe was developed and implemented in COMSOL in order to validate the zero-net-flux approach, using measurement of extracellular brain glucose levels as a well-explored example system against which to compare the model. Results from the model accurately reproduced and predicted results from *in vivo* experiments. Importantly, the model predicts that the time for an implanted probe to achieve steady-state equilibrium with the surrounding extracellular fluid is on the order of one to two minutes, supporting the validity of this technique for quantitative measurement of *in vivo* neurochemistry. **Key Words:** Glucose—Extracellular fluid—Microdialysis—Zero-Net-Flux—Model—Mass transport.

# List of Symbols

| | | |
|---|---|---|
| ρ | (kg/m³) | Density of perfusate |

ρ   (kg/m$^3$) Density of perfusate

$v$   (m/s) Average velocity

$\vec{v}$   (m/s) Velocity vector

$\nabla$   (1/m) Gradient

t   (s) Time

p   (Pa) Pressure

μ   (Pa s) Dynamic viscosity

f   (N) Body forces

Re   (unitless) Reynolds number

$w_m$   (m) Width of the membrane

$w_b$   (m) Width of brain section

$w_a$   (m) Width of the annulus

D   (m$^2$/s) Diffusion coefficient of glucose

c   (mol) Concentration of glucose

$D_B$   (m$^2$/s) Diffusivity constant for glucose through brain

$D_{ECF}$   (m$^2$/s) Diffusivity constant for free diffusion of glucose through extracellular fluid

$\Phi_e$   (unitless) Volume fraction of extracellular fluid in brain

λ   (unitless) Tortuosity factor

## 1. Introduction

Our ability to understand and measure *in vivo* neurochemistry has been greatly enhanced, over the past 25 years or so, by the use of cerebral microdialysis: building on early methods for indirect calculation of extracellular concentration (Lonnroth *et al.* 1987) and determination of appropriate timeframes for measurement to avoid interference from e.g. gliosis (Benveniste & Diemer 1987, Benveniste *et al.* 1987), and moving relatively rapidly from measurements in anesthetized animals to the use of awake, unrestrained rats (Fellows *et al.* 1992) as a gold-standard system, microdialysis has been used to sample and analyze a variety of small molecules including neurotransmitters, metabolites, cytokines, and small peptides in the brain's extracellular fluid (ECF) (Sandberg *et al.* 1986, Wade *et al.* 1987, Kuhr *et al.* 1988, Yamaguchi *et al.* 1990, Fellows et al. 1992, Orosco *et al.* 1995, Shuaib *et al.* 1995, McNay & Gold 1999, McNay *et al.* 2000, McNay & Gold 2001, McNay *et al.* 2001, Cirrito *et al.* 2003, Kino *et al.* 2004, De Bundel *et al.* 2009). Brain metabolism, in particular, has been commonly studied via the use of *in vivo* microdialysis to monitor glucose and other metabolites (lactate, pyruvate, etc.) in specific brain regions during a variety of manipulations including cognitive demand, exogenous pharmaceutical manipulation, hypoglycemia, tactile stimulation, and so on, as well as in some cases being used as a tool to deliver glucose to a targeted brain region (Benveniste et al. 1987, Heyes *et al.* 1991, Fellows et al. 1992, Vahabzadeh *et al.* 1995, Forsyth 1996, Borg *et al.* 1997, Ragozzino *et al.* 1998, Goodman *et al.* 1999, McNay & Gold 1999, McNay et al. 2000, McNay & Gold 2001, McNay et al. 2001, Stefani & Gold 2001, Abi-Saab *et al.* 2002b, McNay & Sherwin 2004a, McNay & Sherwin 2004b, Canal *et al.* 2005, McNay *et al.* 2006a, De Bundel et al. 2009, Zielke *et al.* 2009, Meierhans *et al.* 2010). Of course, brain function is critically dependent on glucose as a source of energy, and several of the above studies have shown – consistent with results obtained using e.g. the 2-deoxyglucose method, or PET and fMRI measurements - that there are localized drops in extracellular glucose concentration in areas of the brain being used, frequently accompanied by increases in extracellular lactate and hence assumed to reflect increased local glycolytic metabolism.  Acute drops in local glucose concentration appear to reflect increased local neural activity, with the magnitude of glucose decrease correlating with, for example, the complexity of a cognitive challenge (McNay *et al*. 2000); conversely, extracellular lactate is commonly elevated  (Goodman *et al*. 1999, Abi-Saab *et al*. 2002a, McNay *et al*. 2006b, Zielke *et al*. 2009, McNay *et al*. 2010), consistent with the suggestion that local glycolysis (specifically, as opposed to oxidative metabolism) is upregulated to support increased local neural activity, an example of neurometabolic coupling that has been observed and studied for well over a hundred years (Roy & Sherrington 1890, Sokoloff 1977, Sokoloff *et al*. 1977, Mercer & Dunham 1981, Mori *et al*. 1990, Sokoloff 1992, Pellerin & Magistretti 1994, Takahashi *et al*. 1995, Chih *et al*. 2001, Brennan *et al*. 2006).

Although accuracy of glucose measurements using microdialysis has improved both with advances in probe design and with superior perfusate solutions that more accurately mimic the ECF composition and hence avoid confounding measurements by altering local ion concentrations (McNay & Sherwin 2004b), a potential drawback to the use of microdialysis, as with many steady-state fluid flow systems, is that the probe fluid may not reach material equilibrium with the brain fluid.  Therefore, for quantitative measurements of brain glucose, concentration is commonly inferred by interpolation to a curve obtained by perfusing a microdialysis probe with a range of different glucose concentrations and recording the outlet concentration, the method of zero-net-flux (ZNF) (Lonnroth *et al*. 1987): when the difference between the inlet and outlet concentrations are plotted against inlet concentration a linear relationship is observed. The point at which the line of best-fit crosses the x-axis is the concentration of the brain ECF, because it is at this point that inlet and outlet concentrations are identical – the point of ZNF. In theory, ZNF is an accurate and reliable method to calculate brain concentration from samples taken with relatively high temporal resolution. However, the ZNF method is based on some important transport assumptions, most notably being that the system is at steady-state for the duration of the experiment. This assumption is challenging to prove experimentally *in vivo*.

    A potential alternative to direct experimental verification of steady-state is the use of a mathematical model of the brain-probe system, within which the relevant parameters can be defined. If the model both (i) accurately simulates experimentally determined ECF glucose values and (ii) shows that the system is in steady-state, then it is reasonable to accept the assumptions of ZNF as valid, providing confidence in ZNF-derived concentrations.

    Several attempts have been made to accurately model the microdialysis system, both *in vitro* and *in vivo*. For example, Bungay *et al*. (2011) constructed a 2 dimensional model that is capable of accommodating both *in vitro* and *in vivo* methods. However, this model used equations that were not based on first principles and assumed a linear concentration profile. Wang *et al*. (2008) used a mathematical representation to model some aspects of fluid flow in microdialysis. The focus of that model was on the temporal resolution of *in vitro* measurements, using a segmented flow approach to model a transient environment and investigate sample diffusion within tubing after leaving the probe; the model was of a different probe geometry and assumed constant diffusion throughout the system. Other models have modeled e.g. pharmacokinetic uses of microdialysis (Morrison *et al*. 1991) or have looked at uses in non-brain

tissues. In contrast, the present model attempts to study the effects of changes in the brain environment on microdialysis probe measurements, and to determine the conditions under which steady-state measurement may be achieved, with inputs to the model being set either by experimental manipulation (e.g. input concentration) or derived from literature values, without additional adjustment; results were consistent in both a full three-dimensional model system and a two-dimensional slice of that model using the same values and approach.

## 2. Materials and Methods

A mathematical representation of microdialysis in the brain was constructed to model mass transfer of glucose, including both (i) by diffusion (through the perfusate fluid, membrane, and immediately adjacent brain), and (ii) by fluid flow within the perfusate.

The calculations used in the model are based on two governing differential equations: Navier-Stokes (Equation 1) (Bird *et al*. 1960) and the conservation of mass (Equation 4) (Bird *et al*. 1960). The Navier-Stokes equation describes the motion of fluid in three-dimensional space where flow rate may vary over time. The first term on the left-hand-side of Equation 1 represents variations in fluid velocity with respect to time; for example, when flow in a system is started. The second term on the left-hand-side represents how velocity may vary at different locations in the system. The first and second terms on the right hand side of Equation 1 represent the pressure gradient and stress, respectively. The third term on the right hand side is external forces such as gravity and electromagnetism; these can be assumed to be negligible in the context of a brain-probe system. In the brain-probe system, the Navier-Stokes equation is used only to model the flow of the perfusate; the membrane and brain tissue are assumed to have no bulk flow of fluid.

$$\rho \left( \frac{d\vec{v}}{dt} + \vec{v} \cdot \nabla \vec{v} \right) = -\nabla p + \mu \nabla^2 \vec{v} + f \quad (1)$$

Where ρ is perfusate density

        *v* is the velocity of the perfusate

        ∇ is the gradient, or differential direction

        p is pressure

        μ is perfusate viscosity

        *f* represents external body forces

The Reynolds number (Equation 2) (Bird *et al*. 1960) is used here to determine the type of flow found within the outer annulus of the microdialysis probe. The Reynolds number, being on the order of 0.02, tells us that the flow has a parabolic profile and can be classified as creeping flow (Stokes flow) (Re«1).

$$Re = \frac{\rho w_a v}{\mu} \quad (2)$$

where $w_a$ is the width of the annulus

In creeping flow (Stokes, 1851), the inertial terms (pressure gradient and spatial acceleration) are negligible compared to temporal acceleration and stress forces. This allows simplification of Equation 1 to Equation 3 (Bird *et al*. 1960), which will save computing power while still providing accurate results.

$$\rho \frac{d\vec{v}}{dt} = \mu \nabla^2 \vec{v} \qquad (3)$$

Equation 3 is used to describe the flow of perfusate through the microdialysis probe.

The conservation of mass equation (Equation 4) describes the transport of material by both convection and diffusion. It is used to describe the diffusion of glucose from the brain, through the membrane and into the perfusate as well as the convection of the glucose due to fluid flow within the perfusate. The first term on the left-hand-side represents an unsteady-state change in concentration, such as experienced during startup of flow in the system. The second term on the left-hand-side represents diffusion due to a concentration gradient, and the third term represents transport due to convection. Use of equation 4 assumes that the concentration of the solute of interest (glucose) is small with respect to the concentration of other components. In this case, and in the case of essentially all microdialysis experiments, the concentration of water in the system is more than two orders of magnitude larger than that of glucose. A further assumption is that there is no bulk flow of fluid through the brain or membrane. Equation 4 neglects the presence of convection in the brain and membrane and would therefore produce inaccurate results if convection were present. If other solutes (e.g. sodium or other ions) were experiencing significant net transport across the membrane, the concentration difference would lead to an osmotic pressure gradient, which would cause a bulk flow (convection). This is minimized or prevented, in *in vivo* microdialysis studies, by the use of an accurate recipe for artificial ECF (aECF) as the perfusate (McNay and Sherwin, 2004), which matches ECF concentrations of ions present at significant levels. This is an important limitation to the present work: studies using nonphysiological perfusates, such as Ringer's solution or even artificial CSF as opposed to aECF (McNay and Sherwin, 2004) may not meet the requirements for the present approach to be applicable in validating use of the ZNF approach. Osmotic pressure could be incorporated into the model; however, we are confident that in general the concentration of other species (neurotransmitters, peptides, metabolites, etc.) in the extracellular fluid is sufficiently small compared to that of water that osmotic pressure is negligible.

$$\frac{dc}{dt} + \nabla \cdot (-D\nabla c) = -\vec{v} \cdot \nabla c \tag{4}$$

Where c is the concentration of the analyte

*D* is the diffusion coefficient of the analyte in the appropriate medium

*2.1. Model Geometry*

The model is designed to simulate a representative modern microdialysis probe; specifically, a CMA 12 microdialysis probe (CMA/Microdialysis, Sweden) with a 500 μm outer diameter and a 3 mm length of polycarbonate membrane (20 kDa cutoff) at the tip placed within a section of brain, to allow comparison of results obtained using the model with our previous *in vivo* glucose microdialysis measurements obtained using this probe type. The probe shaft has a concentric geometry with a stainless steel casing. Perfusate flows through the center of an inner cannula from the top and enters into the outer annulus by means of two bore holes near the base of the inner wall. Although CMA reports membrane thickness to be 25 μm, the membrane material will swell when immersed in fluid; Rosenbloom *et al.* (2005) measured a fully-hydrated polycarbonate membrane to have a thickness of 40 μm using a scanning electron microscope. The brain section being modeled has a thickness of 50 μm, chosen as the maximum distance between a brain location and a capillary (Masamoto *et al.* 2007, Wang *et al.* 2008). The capillary itself is not being modeled, but the concentration of glucose on the outer boundary of the capillary is assumed to be constant due to its proximity to a glucose source.

The brain-probe system was described using the finite element method (FEM) program, COMSOL 4.2 (COMSOL, Inc., Burlington, MA). FEM is used to apply and approximate solutions to Equations 3 and 4 because a closed-form solution to the differential equations is not possible for time-varying problems with complex geometry.

Both a full 3D model of the system (including the bore holes at the tip; here referred to as the 'asymmetric' model) and a 3D axisymmetric model (one that assumes the system is radially symmetric, and hence treats the bore holes as a bore annulus) were constructed, in order to assess the influence of the asymmetric bore holes on fluid flow and concentration profiles. Figure 1 shows the differences between the asymmetric and axisymmetric models. The two models produced identical results under the base-case conditions found in Table 1, and we used the axisymmetric model for the subsequent evaluation of ZNF.

It should be noted that under the conditions in Table 1, the two models in Figure 1 are interchangeable. The only difference in the two models is that in the asymmetric model (A), perfusate enters the inner cannula as the top and flows into the outer annulus by means of two bore holes near the base of the probe. In the axisymmetric model (B), perfusate enters at the base of the outer annulus. It was demonstrated that the difference in flow pattern has no significant effect on the outlet glucose concentration. The axisymmetric model produces results as accurate as the asymmetric model while shortening run times and requiring less computing power.

The mesh used for the model is physics controlled and uses a "Normal" predefined size. This means that each of the above transport equations is solved at 5597 points on the model; an overall solution is then interpolated between the points. The model, when run on a computing cluster (11 Dell PowerEdge Servers with 16-192 GB of RAM per server and a total of 128 CPU cores) can solve for four different inlet glucose concentrations in less than twelve minutes, compared to more than 240 min using the asymmetric model.

*2.2    Model Description*

Equations 3 and 4, and finite element modeling, allow computation of systems that are well defined by continuum mechanics. In order to describe non-homogeneous materials such as the brain and the microdialysis membrane within such a model, these elements are described by a homogeneous approximation. This is done by replacing the diffusivity of glucose with an *effective* diffusivity which accounts for the presence of effectively impermeable materials within the system. Brain effective diffusivity was obtained using Equation 5, which models the movement of glucose through the brain in terms of tortuosity: the more tortuous a material is the longer the path a molecule has to traverse to get from one point to another. The path of diffusion of glucose from a blood vessel to a probe is impeded both physically and chemically by brain cells. Not all of the glucose that leaves a capillary makes it to the probe; some of the glucose is absorbed by brain cells. And the glucose that does make it to the probe may not have diffused there in a straight line, but instead moved around the brain cells. To build these diffusion inhibition factors into the diffusion coefficient, the constant for the free diffusion of glucose in ECF ($D_p$) is multiplied by the volume fraction of extracellular fluid in brain ($\phi_e$) and then divided by the square of the tortuosity factor ($\lambda$), producing Equation 5 (Benveniste *et al.*, 1989) which allows us to model the complex, nonhomogeneous brain as a homogenous medium amenable to computation.

$$D_B = \frac{D_p \phi_e}{\lambda^2} \tag{5}$$

Effective diffusivity in the probe membrane is also described by Equation 5. Effective diffusivity is also available from experimental results for the same probes as described by Buttler *et al*. (1996) and Wisniewski *et al*. (2001) and these are used in preference to calculated values.

Parameters used in this description are summarized in Table 1. Note that the "outer boundary" concentration is the model's "true" concentration of glucose within the brain ECF, the concentration that ZNF is used to assess. One of the strengths of the modeling approach is the ability to independently set this as a parameter and assess whether ZNF experiments within the model predict the known "brain" concentration.

Values for volumetric flow rate, initial perfusate/membrane and brain concentrations, membrane length and outer boundary concentrations in this example are based on those used by McNay and Sherwin (2004), as that is the experiment we use (below) to provide an *in vivo* dataset to confirm accurate simulation; each of these variables can be altered within the model to examine the impact of such experimenter-controlled variables. The width of the membrane is discussed above (Rosenbloom et al, 2005). Density, viscosity and perfusate diffusivity were calculated using the PRSV equation of state at 38.2°C (Aspen Technology, Inc., Burlington, MA). While, historically, the volume fraction of the brain has been taken as 0.2 (see e.g. Benveniste, 1989), somewhat larger values [0.3 - 0.4] have been reported to best fit data obtained in specific studies (Dykstra et al, 1992). None of the results reported here are significantly affected by altering the void fraction used in the model within the range 0.2 - 0.4. We therefore chose to use a volume fraction of 0.35. The tortuosity factor was set at 1.6 (Benveniste, 1989). The membrane diffusivity was derived from characteristic membrane data collected by Buttler *et al*. (1996) and Wisniewski *et al*. (2001).

All system parameters are assessed at rat-core body temperature (38.2°C, Martin and Papp, 1979) rather than laboratory temperature. The validity of this approach was verified by a separate model based on heat transfer. Figure 2 shows the temperature at the center of the perfusate as it travels from the inlet tubing and down the inner cannula. The temperature does not start at room temperature (22.85°C) because heat flows out of the rat's head and up the inlet perfusate channel. It is apparent that the perfusate reaches a steady-state temperature by the time it reaches the bore holes (dashes, Figure 2). The 0.1°C difference between fluid temperature and brain temperature proved to be insignificant.

*2.3    Data Processing*

Microdialysis sampling records the average glucose concentration at the probe outlet over the sampling period; hence, this is the datum needed to verify ZNF. This can be accomplished by integrating the concentration of glucose over the surface of the outlet boundary, a calculation which COMSOL can compute internally.

## 3.    Results

*3.1.    Fluid flow and concentration gradients produced by the model.*

Figure 3 shows cross sections of fluid flow and concentration profiles within the probe at steady-state, using the parameters above (e.g. inlet concentration of 0.50 mM glc, brain concentration of 1.25 mM). Note that only a 2D cross-section of the probe is shown, for ease of visualization; as the probe is radially symmetric, conditions are identical around the central inlet tube. Fluid motion is fully laminar within the perfusate region (Figure 3a), as hypothesized. Figure 3b shows a gradual concentration gradient both axially and radially of the model, as expected.

Figure 3a shows the velocity profile throughout the probe. Note that the membrane and brain have no color, signifying that there is no net flow in those regions. Dark red areas have the fastest moving fluid. The insert, a graphical representation of the same flow data, shows that fluid near the center of the annulus is traveling faster than the fluid on the boundaries. This discrepancy creates a parabolic flow regime, known as laminar flow and is seen in fluids with a Re < 2100.

Figure 3b shows the concentration gradient throughout the brain-probe system. Dark red represents high glucose concentrations and dark blue areas have low glucose concentrations. The upper and lower concentration limits for a $c_{in}$ of 0.50 mM are 1.25 and 0.50 mM, respectively. Figure 3b shows that the perfusate enters the outer annulus with a concentration of 0.50 mM (which is set) and that perfusate concentration gradually increases as it travels up the probe. Figure 3b shows the instantaneous concentration throughout the probe; as noted above, measured sample concentration will be an average of the fluid concentration leaving the probe.

*3.2. Validation of model predictions against in vivo data*

To confirm that our model accurately simulates *in vivo* reality, we set parameters to match those used by McNay and Sherwin (2004) with experimenter-controlled variation in perfusate glucose concentration (1 µL/min flow rate, 2 µL samples taken 60 minutes after start). The model was then used to calculate a predicted net concentration change against inlet concentration (line, Figure 4) which was compared to experimental data obtained by McNay and Sherwin (diamonds, Figure 4). As can be seen in Figure 4, the model accurately recreates the experimentally-determined point of ZNF, and the model's predicted data describe experimental results with an $r^2$ of 0.90: novel predictions of expected concentration produced by varying values of physical parameters within the model are closely matched by experimental observations following identical changes to those parameters.

Importantly, all model parameters were based either on experimentally set parameters (flow rate, inlet concentration) or independently derived values (diffusion coefficients, viscosity, density), using no "fitting" parameters to adjust the model. We hence conclude that the evidence supports our model as an accurate representation of the brain-probe microdialysis system, allowing us to proceed to ask questions about the validity of assumptions underlying techniques such as ZNF.

Note that validating the model allows for subsequent brain concentrations to be calculated from a single measurement, rather than a complete ZNF series of measurements. Because the slope of the ZNF line is independent of brain concentration, measurement of the outlet concentration for a single known perfusate concentration allows extrapolation to the ZNF point using Equation 6.

$$c_C = c_{\text{perfusate}} - \frac{c_{\text{dialysate}} - c_{\text{perfusate}}}{\text{ZNF slope}} \tag{6}$$

*3.3    Attributes of ZNF demonstrated by the model*

3.3.1. Robustness of ZNF to changes in brain concentration

Using Equation 6, we can extrapolate a single measurement to obtain the actual brain concentration at the time of sampling. This is of obvious importance for quantifying changes in brain concentration with time, but accurate quantification requires the assumption that the slope of the ZNF plot does not vary with brain concentration. Figure 5 shows that this is indeed the case: the slope is a function of experimental and physical parameters and does not vary with changes in brain concentration.

Note that importantly, although we use glucose here as an illustration, the results obtained through use of our model are applicable to measurements for any analyte.

3.3.2. Sensitivity of ZNF to experimental parameters

A sensitivity analysis was run on each of the model parameters to determine the influence of each physical parameter on the resulting perfusate outlet concentration, holding all other factors constant at the values shown in Table 1. This approach was used to determine which factors are most important to the overall result and to what extent uncertainty in physical parameters propagates to results. Each input variable was independently tested at 13 different values over a range of 10% to 1000% of the normal value. Outlet glucose concentration was used as the dependent variable.

Figure 6 shows the results of this analysis. As expected, outlet glucose concentration is significantly sensitive to flow rate. Indeed, this sensitivity forms the basis of the 'zero-flow' method of extrapolation from microdialysis sample concentration to actual brain concentration, commonly used for low-abundance species (Goossens *et al*. 2004), in which flow rate is varied and results extrapolated to a theoretical point of zero flow at which complete equilibration of the perfusate with extracellular fluid would be achieved. As long as flow rate is constant throughout any given experiment, however, accurate results may be obtained from any flow rate. In contrast, a change in perfusate density, diffusivity or viscosity within one order of magnitude has no effect on the outlet glucose concentration.

Results are somewhat sensitive to the volume of the 'brain' compartment, modeled here as an annulus around the probe. The model used here uses a width for this compartment of 50 μm of brain, because that should be approximately the upper limit for distance between a probe membrane and a source of glucose external to the model (i.e. a capillary) (Masamoto *et al*. 2007), and it is therefore unlikely that the error level produced by a 1000% increase in this parameter is biologically plausible. Sample concentration is also somewhat sensitive to membrane diffusivity: if membrane diffusivity varies by an order of magnitude, it can shift the results by as much as 20%. We were able to obtain multiple consistent literature values for the diffusivity of the membrane in our model (Buttler *et al*. 1996, Wisniewski *et al*. 2001), and the close match of our model's results to *in vivo* data suggests that these were accurate, but this sensitivity should be borne in mind if attempting to vary membrane material type (e.g. if altering material and/or pore size, such as when sampling relatively large species such as peptides). One variable not included in the model is any effect of glial scarring or acute traumatic damage from probe insertion, which has previously been modeled and shown to be potentially important in the context of dopamine microdialysis (Bungay *et al*. 2003). Glucose measurements using microdialysis are affected by such variables on a timescale of days (Benveniste & Diemer 1987, Benveniste et al. 1987), as well as by other variables such as changes in extracellular tortuosity across the lifespan (McNay & Gold 1999), but are consistent with values from other techniques when taken within 24h of initial probe insertion. The fact that our model for glucose matches *in vivo* data without including this variable may, perhaps, reflect tighter sequestration and regulation of neurotransmitters such as dopamine than glucose in healthy brain tissue.

*3.4. Time to achieve steady-state for ZNF*

For calculation of brain concentration using ZNF to be accurate, the system has to be in steady-state. We hence modeled how long it would take for steady-state to be achieved after starting perfusion, and in response to an acute alteration in brain concentration (such as that seen, for example, in glucose when cognitive activity increases and hence demand for glucose is higher, causing a sudden dip in local brain glucose concentration; tracking of acute decreases in local ECF glucose as a marker for local neural activity during cognitive testing was first reported by McNay *et al*. (2000) and has been replicated by several groups in varying conditions (e.g. (Rex *et al*. 2009, Newman *et al*. 2011)).

3.4.1. Time to achieve initial steady-state

Figure 7 shows the modeled outlet concentration obtained, across time, after starting microdialysis perfusion with a perfusate containing one of four glucose concentrations and a brain glucose concentration of 1.25 mM. Steady-state occurs when outlet concentration does not vary further with time, producing a horizontal plot; it can be seen that each plot asymptotically approaches this condition and steady-state is achieved within a few minutes even for perfusate concentrations quite different from the brain concentration. Table 2 summarizes how long it takes the system to reach several degrees of close approximation to steady-state, using an inlet concentration of 0.25 mM. Within two minutes of starting perfusion, the system will have effectively reached steady-state. Because this time is small compared to that needed after probe insertion to allow resealing of the blood-brain barrier around the probe (Benveniste *et al*. 1989), steady-state is likely to be achieved, at least at baseline, in all sensibly-designed microdialysis experiments, providing reassurance that the requisite conditions underlying ZNF calculations will be met.

3.4.2 Time to respond to acute changes in brain concentration

In addition to measurements of baseline concentration, a key use for *in vivo* microdialysis is to follow, with good temporal resolution, acute changes in neurochemistry (neurotransmitter release, glucose usage, etc.). Quantification of such changes, though, relies on steady-state for ZNF calculations just as baseline measurements do, something that may require significant time and hence limit the temporal resolution achievable. We thus used the model to ask how fast steady-state would be re-achieved in response to such an acute change; specifically, as an exemplar case, we used the model to determine how long the system takes to adjust to a 0.25 mM drop in brain concentration, solved over 0 to 750 seconds with sample concentration calculated every second. The model results are shown in Figure 8 and times to steady-state are summarized in Table 3.

It should be noted that the drop in outlet glucose concentration when steady-state is re-achieved is independent of perfusate concentration. In the case of Figure 8, the drop in brain glucose concentration was 0.25 mM while the drop in outlet glucose concentration, for all perfusate concentrations, was 0.061 mM.

Table 3 shows that the closer inlet concentration is to the brain concentration, the quicker the system reaches steady-state. Hence, use of perfusate with appropriate concentrations will improve temporal resolution, but even marked deviation from optimum perfusate composition still results in temporal resolution on the order of 1-2 minutes; it should also be noted that the time for outlet concentration to alter in response to a change in brain concentration includes a 16.1s residence time for perfusate to travel from probe tip to outlet. This timescale matches well with results from several groups suggesting that changes in small molecule concentration can be observed on the order of one minute.

4. **Discussion**

The model described here is a powerful tool in support of microdialysis measurements. It allows independent assessment of assumptions within the ZNF experiments, such as the attainment of steady-state, as well as the testing of situations that are not directly accessible to experiment, such as the impact of a step-change in brain-glucose concentration.

If perfusate, dialysate and environmental concentrations are known, they can be used to determine the diffusivity of a certain analyte through the membrane. The current model assumes that effective diffusivity may be treated as a constant for both the brain and the membrane. Wisniewski *et al*. (2001) and Butler *et al*. (1996) have demonstrated that membrane fouling and structural changes in the brain, respectively, will reduce the effective diffusivity over the course of several days. What is suggested by the COMSOL model is that the rate of fouling is low enough that a constant diffusivity is accurate within the context of a single experiment (~1 hour), while for longer-term changes (~days) the fouling might be important to consider when comparing, for example, multiple measurements taken from a single animal across a multi-day experiment. Future work could incorporate a time-dependent description of effective diffusivity.

It was found that membrane diffusivity is indirectly related to the velocity in the outer annulus. For a slow moving perfusate, a static, no-slip boundary layer is formed on the inside of the membrane. This boundary layer increases the overall mass transfer resistance of the system. As the velocity is increased, the boundary layer becomes thinner which lowers overall resistance and in turn, increases the membrane diffusivity, so that increasing experimental flow rate does not reduce sampling from the brain as much as might otherwise have been expected.

Overall, the present data show that the model created in COMSOL accurately reflects the brain-probe system, with results that match experimental data. Use of the model shows that the assumptions underlying ZNF are likely to be accurate for the majority of *in vivo* microdialysis measurements, and provides guidance both for the design of microdialysis experiments and for the limits of resolution and interpretability likely to exist for data obtained from such experiments, with results supporting *in vivo* studies suggesting that changes in ECF concentration can be measured using microdialysis with temporal resolution on the order of a minute. Models such as that discussed here offer the potential for valuable insights into brain microdialysis sampling, and may be used to investigate problems that are difficult to address in the laboratory, or would require knowledge of experimentally unmeasureable variables.

5. **Acknowledgements**

We would like to thank Dr. James Maneval for his help with all things math. The initial suggestion that mathematical modeling might assist microdialysis experimental design came from Dr. Jennifer McNay. This work was supported in part by NIH R01 DK 077106 (ECM).

**Table 1**

| Variable | Expression | Value | Units |
|---|---|---|---|
| $V$ | Volumetric flow rate | $1.67 \times 10^{-11}$ | m³/s |
| $c_{0P}$ | Initial concentration in perfusate/membrane | 0.50 | mol/m³ |
| $c_{0B}$ | Initial concentration in brain | 1.25 | mol/m³ |
| $h_m$ | Height of membrane | $3.00 \times 10^{-3}$ | m |
| $c_c$ | Outer boundary (brain) concentration | 1.25 | mol/m³ |
| $w_m$ | Thickness of membrane | $4.00 \times 10^{-5}$ | m |
| $w_b$ | Width of brain section | $5.00 \times 10^{-5}$ | m |
| $\rho_p$ | Density of perfusate | $9.90 \times 10^{2}$ | kg/m³ |
| $\eta_p$ | Dynamic viscosity of perfusate | $7.28 \times 10^{-4}$ | kg/m s |
| $D_P$ | Diffusion coefficient of glucose in perfusate | $8.30 \times 10^{-10}$ | m²/s |
| $D_B$ | Diffusion coefficient of glucose in brain | $7.62 \times 10^{-11}$ | m²/s |
| $D_M$ | Diffusion coefficient of glucose in membrane | $3.32 \times 10^{-11}$ | m²/s |

**Table 2**

| Proximity to Steady-State (%) | Time to steady-state (s) |
|---|---|
| 90 | 21 |
| 99 | 43 |
| 99.9 | 67 |
| 99.99 | 90 |
| 99.999 | 113 |

**Table 3**

| Proximity to Steady-State (%) | Time to steady-state (s) | |
|:---:|:---:|:---:|
| | $C_{in}$ = 0.25 mM | $C_{in}$ = 1.00 mM |
| 110 | 18 | 0 |
| 101 | 47 | 32 |
| 100.1 | 71 | 40 |
| 100.01 | 95 | 65 |
| 100.001 | 119 | 89 |

**Figure 1**

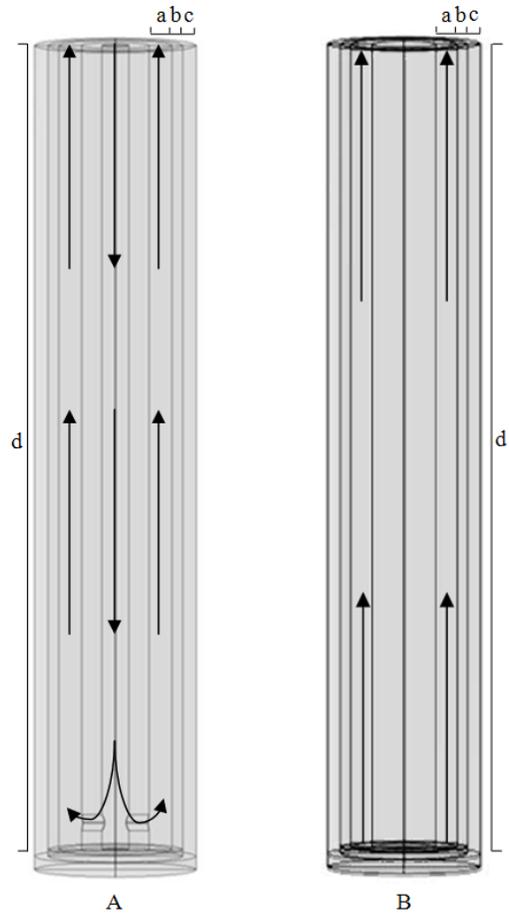

A  B

**Figure 2**

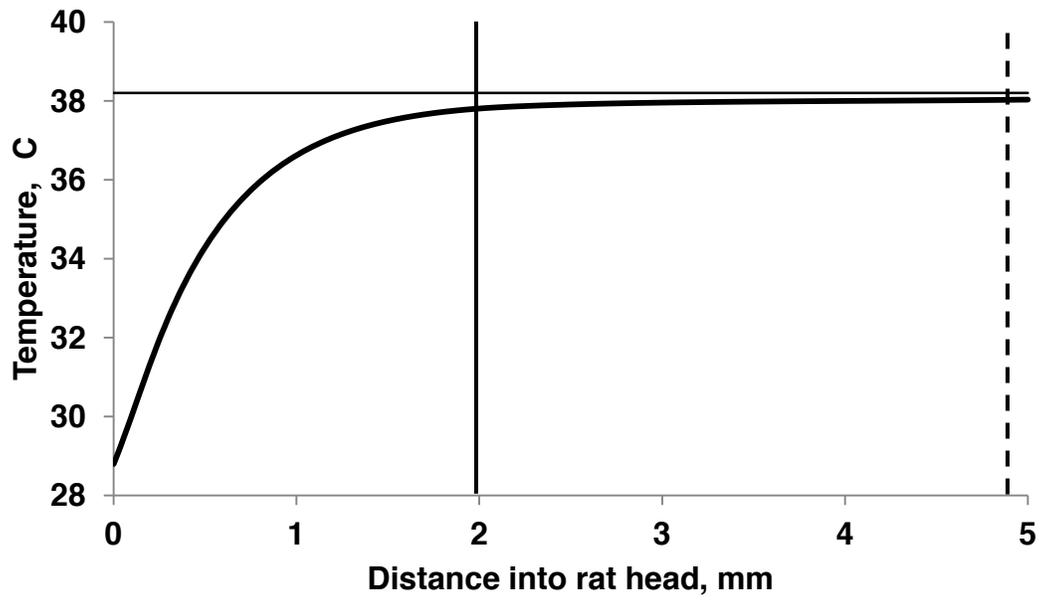

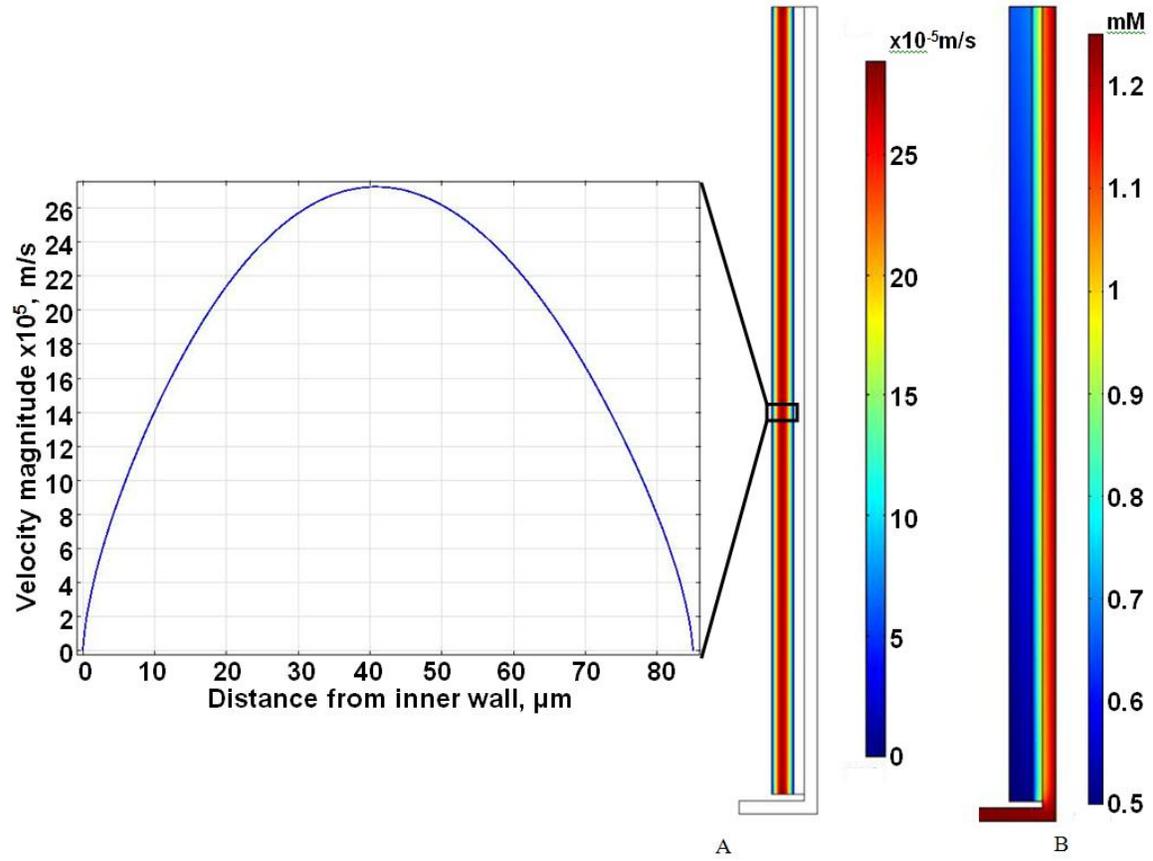

**Figure 4**

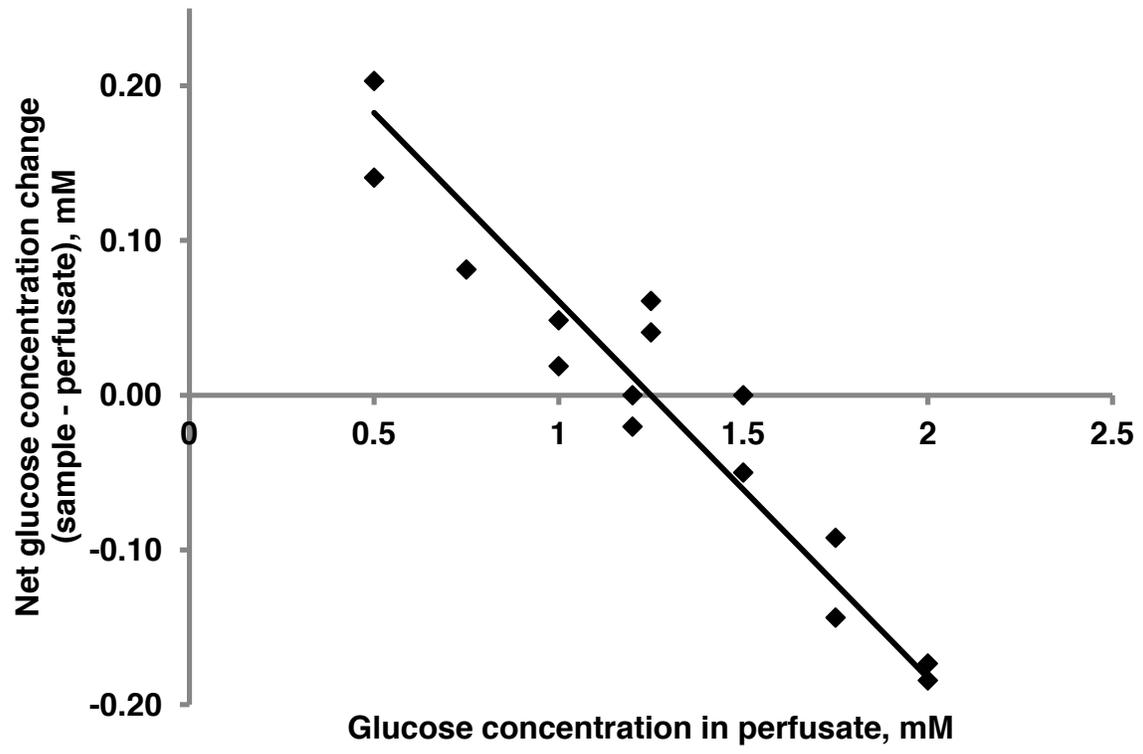

**Figure 5**

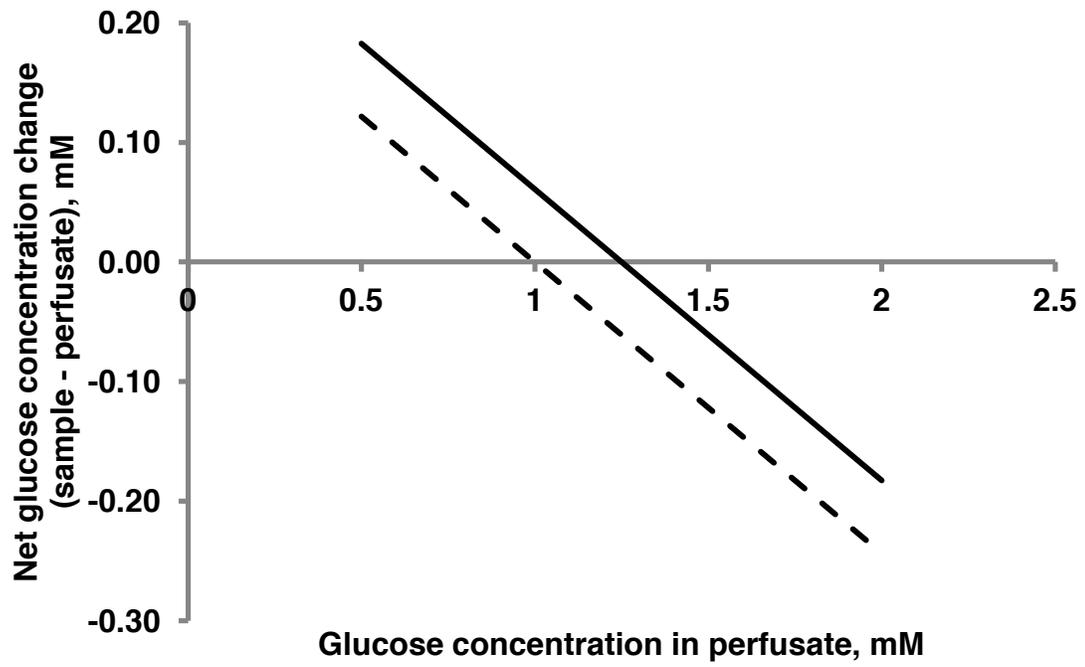

**Figure 6**

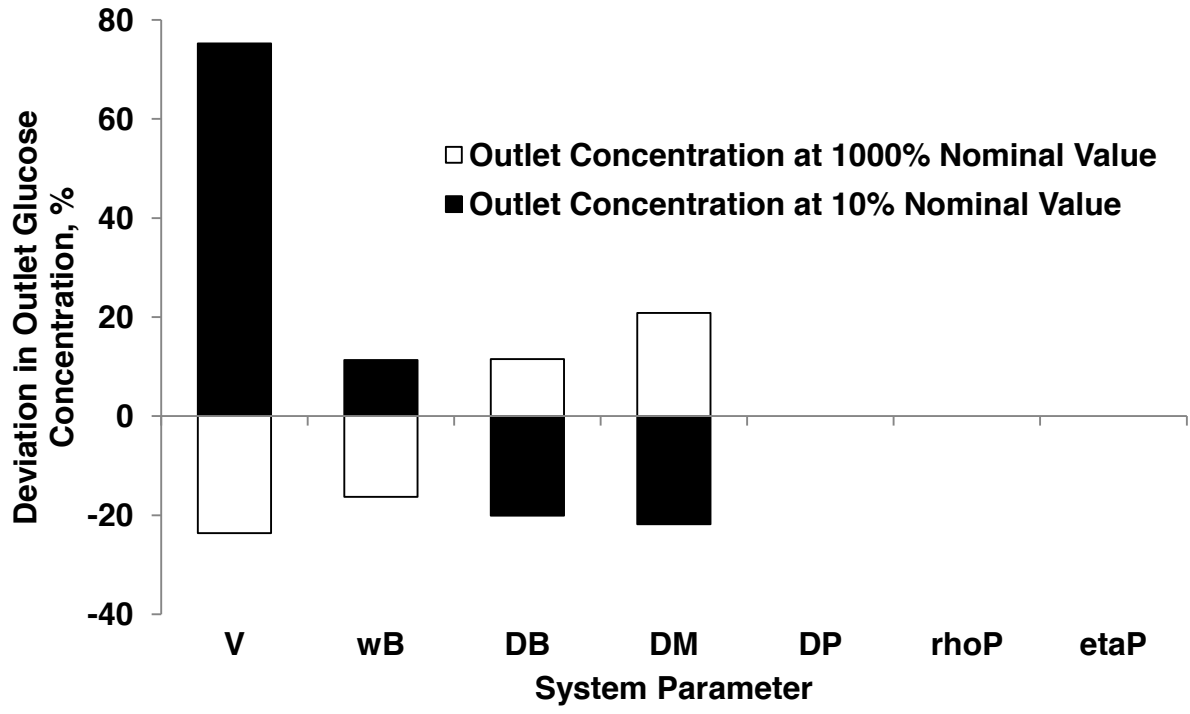

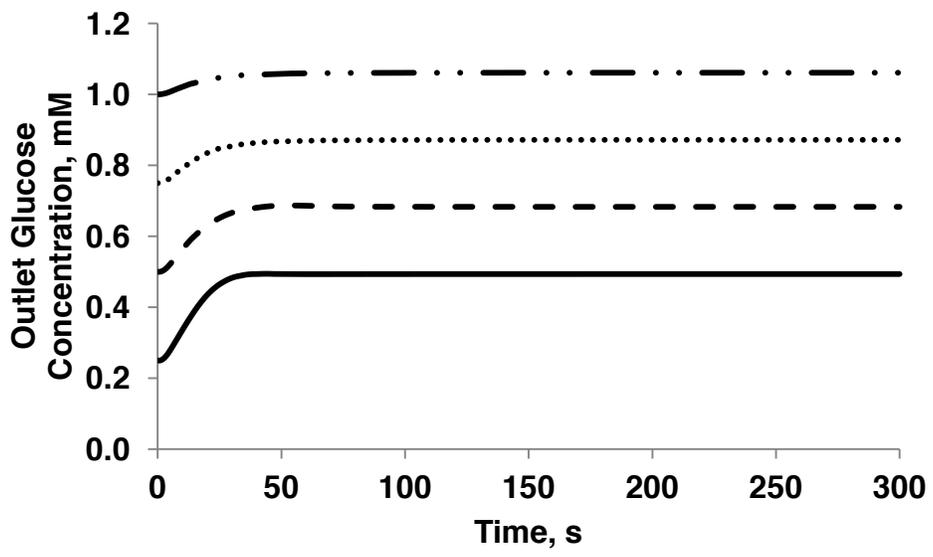

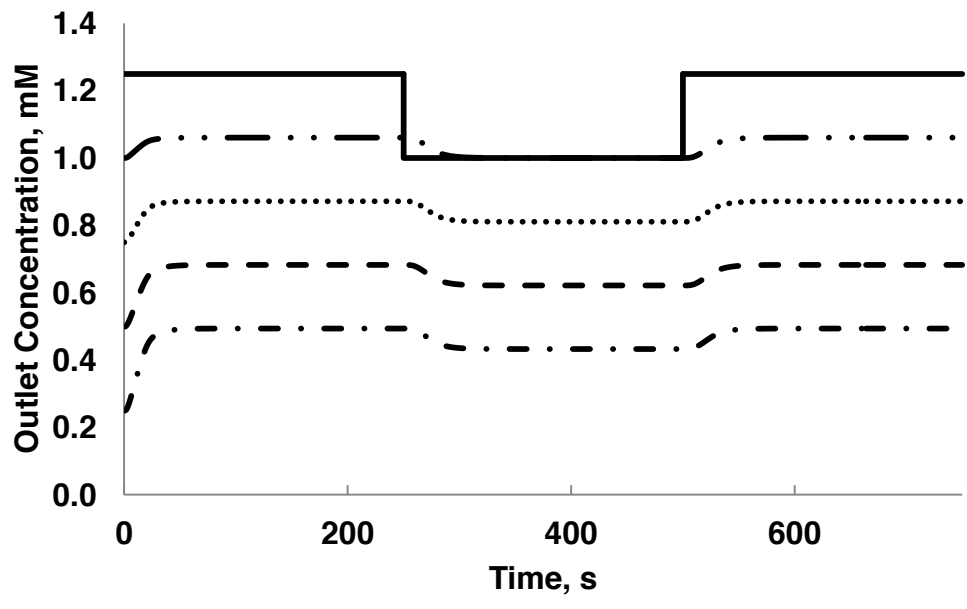